# Optical Birefringence and Domain Structure of the as-Grown Tl$_{2x}$Rb$_{2x(1-x)}$Cd$_2$(SO$_4$)$_3$ Solid Solutions at Phase Transitions

[1]Say A., [2]Sveleba S., [1]Teslyuk I., [1]Martynyuk-Lototska I., [1]Girnyk I. and [1]Vlokh R.

[1]Institute of Physical Optics, 23 Dragomanov St., 79005 Lviv, Ukraine
e-mail: vlokh@ifo.lviv.ua
[2]Electronics Department, Lviv National University, 107 Tarnavski St., 79017 Lviv, Ukraine

## Abstract

It is shown that the as-grown solid solutions Tl$_{2x}$Rb$_{2(1-x)}$Cd$_2$(SO$_4$)$_3$ possess a residual optical birefringence in the cubic phase at room temperature. The existence of both the residual birefringence and the residual domain structure in the compound with $x = 0.8$ at the room temperature, along with the domain walls orientation and the orientation of extinction positions of the neighbouring domains, indicate that the above domain structure corresponds to "forbidden" ferroelastic domains of the phase $P2_12_12_1$. On the basis of measurements for the temperature variation of the birefringence and observations of the domain structure transformations, one can conclude that the phases with the symmetries $P2_13$ and $P2_1$ coexist above $T_{c1}$ in the compounds with $x = 0.7 – 1.0$. The phase transition temperatures $T_{c1}$, $T_{c2}$, and $T_{c3}$ obtained from the temperature birefringence variations agree well with those obtained previously, using the studies of thermal expansion and ultrasonic wave velocities.

**Key words**: optical birefringence, phase transitions, langbeinite crystals, Tl$_{2x}$Rb$_{2(1-x)}$Cd$_2$(SO$_4$)$_3$ solid solutions.

## Introduction

Crystals belonging to langbeinite family (general chemical formula $M_2^+M_2^{++}(LO_4)_3$, where $M_2^+$ and $M_2^{++}$ are respectively monovalent and bivalent metal ions and L stands for S, Se or P) are attractive at least from the viewpoint of their complicated sequence of phase transitions [1]. Let us remind of the main peculiarities of those transitions. The sulphur langbeinites may be divided into following three groups:

(1) crystals that possess just one ferroelastic phase transition, with the symmetry change $P2_13\ F\ P2_12_12_1$;

(2) crystals that manifest ferroelectric-ferroelastic and pure ferroelastic phase transitions, with the changes of symmetry $P2_13\ F\ P2_1\ F\ P1\ F\ P2_12_12_1$;

(3) crystals that do not reveal structural phase transitions at the atmospheric pressure.

We should notice that, although the phase transitions in the langbeinites have been intensively studied, the interest to these crystals is still not reducing. So, the phase diagrams for the mixed crystals of this family have been constructed for a better understanding of the nature of phase transitions in the langbeinites. For example, as reported by *Gustafsson J.C.M. et al* [2] and *Trubach I.G. et al* [3], new mixed phosphate langbeinites Rb$_2$YbTi(PO$_4$)$_3$, Rb$_2$Yb$_{0.32}$Ti$_{1.68}$(PO$_4$)$_3$ and Rb$_2$FeZr(PO$_4$)$_3$ have been synthesized and characterized structurally. Due to our recent studies [4,5], the isolated point of second-order phase transition has been revealed on the line of first-order phase transitions in $K_2Cd_{2x}Mn_{2(1-x)}(SO_4)_3$ crystals, which belong to the group (1) due to their simple sequence of phase transitions. We have also presented [6] the results of studies for $x,T$-phase diagram for $Tl_{2x}Rb_{2(1-x)}Cd_2(SO_4)_3$ crystals, which correspond to the group (2). This phase diagram has been obtained from the measurements of thermal expansion and ultrasonic wave velocity. It has been found in the acoustic wave velocity experiments that the acoustic echo disappears at the temperatures $T = 140\,K$ and $T = 223\,K$ for the compounds with $x = 0.6$ and $0.8$, respectively. These facts, along with a specific behaviour of thermal expansion at these temperatures, would incline one to the conclusion that the additional phase with the symmetry of $R3$ exists in these compounds below the temperatures mentioned above. It has been assumed that the sequence of phase transition in $Tl_{2x}Rb_{2(1-x)}Cd_2(SO_4)_3$ crystals with $x = 0.6$, 0.8 and 1.0 is [$P2_13\ F\ R3\ F\ P2_1\ F\ P1\ F\ P2_12_12_1$], while we have the sequence $P2_13\ F\ P2_1\ F\ P1\ F\ P2_12_12_1$ for the case of $0 \le x \le 0.5$. However, a trigonal symmetry of this phase ($R3$) follows just from the group-theoretical analysis [7] and our observations [8] of the domain structure configuration in the pure $Rb_2Cd_2(SO_4)_3$ crystals, since the latter is typical for the phase transitions with the symmetry change of $P2_13\ F\ R3$. It is necessary to note that the domain configuration in the supposed phase with the symmetry $R3$, occurring in the pure $Rb_2Cd_2(SO_4)_3$ crystals, can be also peculiar for diffuse phase transition with the change of symmetry $P2_13 \Leftrightarrow P2_1$. Hence, the nature of the discussed intermediate phase (or region) in $Tl_{2x}Rb_{2(1-x)}Cd_2(SO_4)_3$ crystals is not yet solved for certain.

It is well known that investigation of temperature dependences of optical birefringence represents one of the most sensitive techniques for determining phase transition temperatures. When applying this method to the langbeinite crystals, which are cubic in their paraelectric-paraelastic phase, one has to determine precisely the temperature points where the birefringence appears or disappears at cooling or heating (i.e., the temperatures of the phase transitions). On the other side, one can derive the symmetry changes at the phase transitions on the basis of information about the domain structure configuration. As a result, the present paper is devoted to studies for temperature variations of the birefringence and polarization-microscopic observations of the domain structure in the solid solutions of $Tl_{2x}Rb_{2(1-x)}Cd_2(SO_4)_3$.

**Experimental**

$Tl_{2x}Rb_{2(1-x)}Cd_2(SO_4)_3$ crystals with $x = 0.1 - 0.9$ ($\Delta x = 0.1$) were grown at 358 K from aqueous solution of [$Rb_{2(1-x)}SO_4$, $Tl_{2x}SO_4$] and $CdSO_4$ salts mixed in the molar ratio 1:2. Single crystals of satisfactory optical quality were obtained after two-month growing process was finished. All the crystals were characterized with a well-defined habit, which is typical for the cubic langbeinites (see, e.g., [9,10]). One compound (that corresponding to $x = 0.9$) was obtained in the shape of plates. The dimensions of crystal pyramids were $8 \times 8 \times 8\,mm^3$. The optical microscopic observations performed at the room temperature revealed a remaining birefringence in the cubic phase of these crystals. Its existence may be caused by a non-uniform distribution of Tl and Rb ions by throughout their proper positions in the lattice. A similar effect has once been observed for the as-grown $K_2Cd_{2x}Mn_{2(1-x)}(SO_4)_3$ solid solutions [11]. This non-uniform distribution of Tl and Rb ions could evidently lead to appearance of mechanical strains.

The residual birefringence was measured on $<111>$-oriented samples, using the standard polarization-microscopic method. The concentration dependence of the birefringence is presented in Fig. 1. It is clearly seen that the residual birefringence increases sharply when $x$ approaches 0.9 and is equal to $\Delta n = 6 \times 10^{-4}$ exactly for the solid solution with $x = 0.9$. Such a large size of the birefringence is quite unusual for the case of residual, built-up effect that normally occurs in crystals with moderate piezooptic coefficients (the piezooptic coefficients for the langbeinites are of the order of a few $pm^2/N$ [12]). This means that any further studies of $Tl_{2x}Rb_{2(1-x)}Cd_2(SO_4)_3$ crystals with $x = 0.9$ would not be transparent until the reasons for the appearance of this birefringence are ascertained.

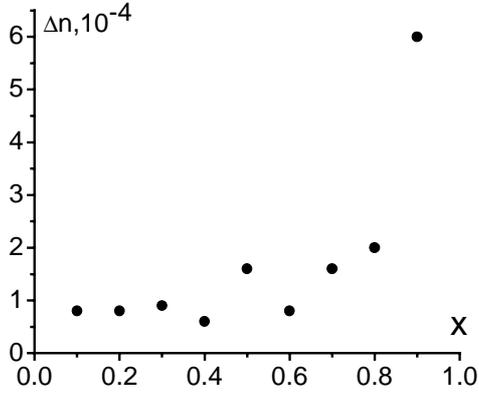

**Fig. 1.** Concentration dependence of the residual birefringence at the room temperature for the as-grown $Tl_{2x}Rb_{2(1-x)}Cd_2(SO_4)_3$ solid solutions.

We studied the temperature dependence of the birefringence with the Senarmont technique in the heating run. In order to cool the samples down to the liquid-nitrogen temperature, a cooling cell was used with the possibility of temperature stabilization not worse than 0.1 K. The samples were oriented according to the Senarmont compensation position in each phase separately. The as-grown samples with the shape of rectangular plates were prepared, with their faces being perpendicular to $<111>$ or $<100>$ crystallographic directions. The domain structure was observed with the aid of Karl-Zeiss polarization microscope.

**Results and discussion**

The temperature dependences of the birefringence for $Tl_{2x}Rb_{2(1-x)}Cd_2(SO_4)_3$ solid solutions with $x=0.1-0.6$ are presented in Fig. 2. The dependences prove to be quite informative for the purpose of determining the phase transition temperatures. In particular, the low-temperature first-order phase transition into the ferroelastic phase is clearly visible for the most of compounds as an abrupt jump of the birefringence. For the compounds with $x=0.7$ and $0.8$, the birefringence does not decrease at $T_{c1}$ down to the residual effect value measured at the room temperature, but gradually changes in the heating run (see Fig. 3). This implies that the optical anisotropy in these crystals slowly decreases with temperature increasing above $T_{c1}$.

Following from the obtained results, one can determine the phase transition temperatures $T_{c1}$, $T_{c2}$ and $T_{c3}$ for the compounds with $x = 0.1-0.8$. Those temperatures are collected in Table 1.

**Table 1.** Temperatures of phase transitions in the as-grown $Tl_{2x}Rb_{2(1-x)}Cd_2(SO_4)_3$ solid solutions.

| Concentration $x$ | $T_{c1}, K$ ($P2_13 \Leftrightarrow P2_1$) | $T_{c2}, K$ ($P2_1 \Leftrightarrow P1$) | $T_{c3}, K$ ($P1 \Leftrightarrow P2_12_12_1$) |
|---|---|---|---|
| 0.1 | 128 | 124 | 101 |
| 0.2 | 127 | 113 | 98 |
| 0.3 | 133 | 119 | 102 |
| 0.4 | 126 | 112 | 97 |
| 0.5 | 125 | 115 | 95 |
| 0.6 | ~117 | 92 | 80 |
| 0.7 | ~124 | 106 | 97 |
| 0.8 | ~127 | 109 | 88 |

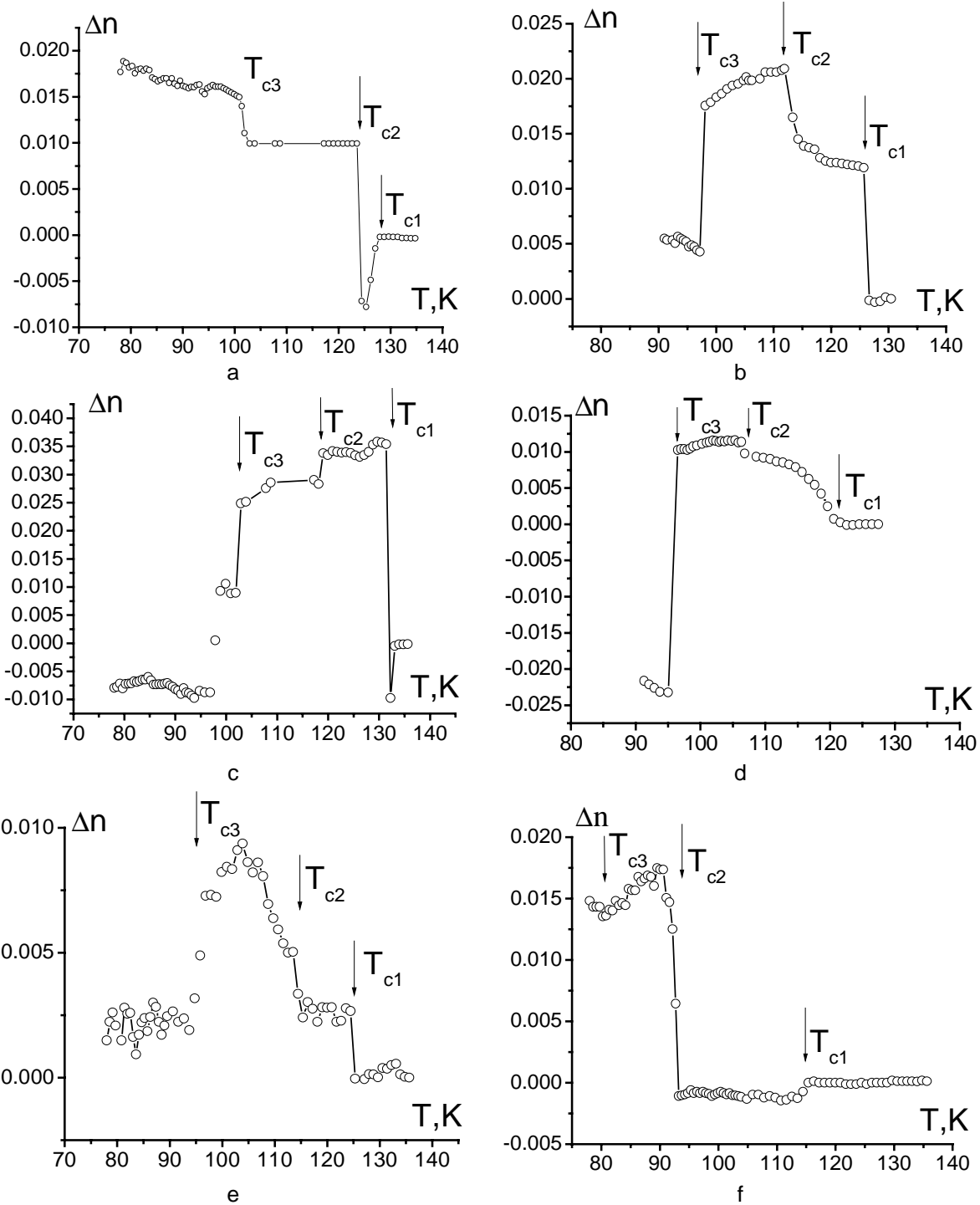

**Fig. 2.** Temperature dependences of the birefringence for $Tl_{2x}Rb_{2(1-x)}Cd_2(SO_4)_3$ solid solutions: (a) $x = 0.1$, (b) $x = 0.2$, (c) $x = 0.3$, (d) $x = 0.4$, (e) $x = 0.5$ and (f) $x = 0.6$ (<111> crystal plates, the light wavelength $\lambda = 632.8$ nm).

Let us now compare our present phase transition points with those obtained in the earlier studies for thermal expansion and ultrasonic wave velocities [6]. One can see that a good agreement with our previous results takes place for the solid solutions with $x = 0.1 - 0.6$. The temperatures of phase transitions differ only due to a hysteresis of the first-order phase transitions (the ultrasonic wave velocities and the thermal expansion have been measured at cooling, while the birefringence data refer to the heating run). As we will show below, the discrepancy regarding the temperature $T_{c1}$ for $x$=0.7 and 0.8 is associated with the coexistence of phases, which is typical for the langbeinites (see, e.g., [13,14]). The corresponding phase diagram is depicted in Fig. 4.

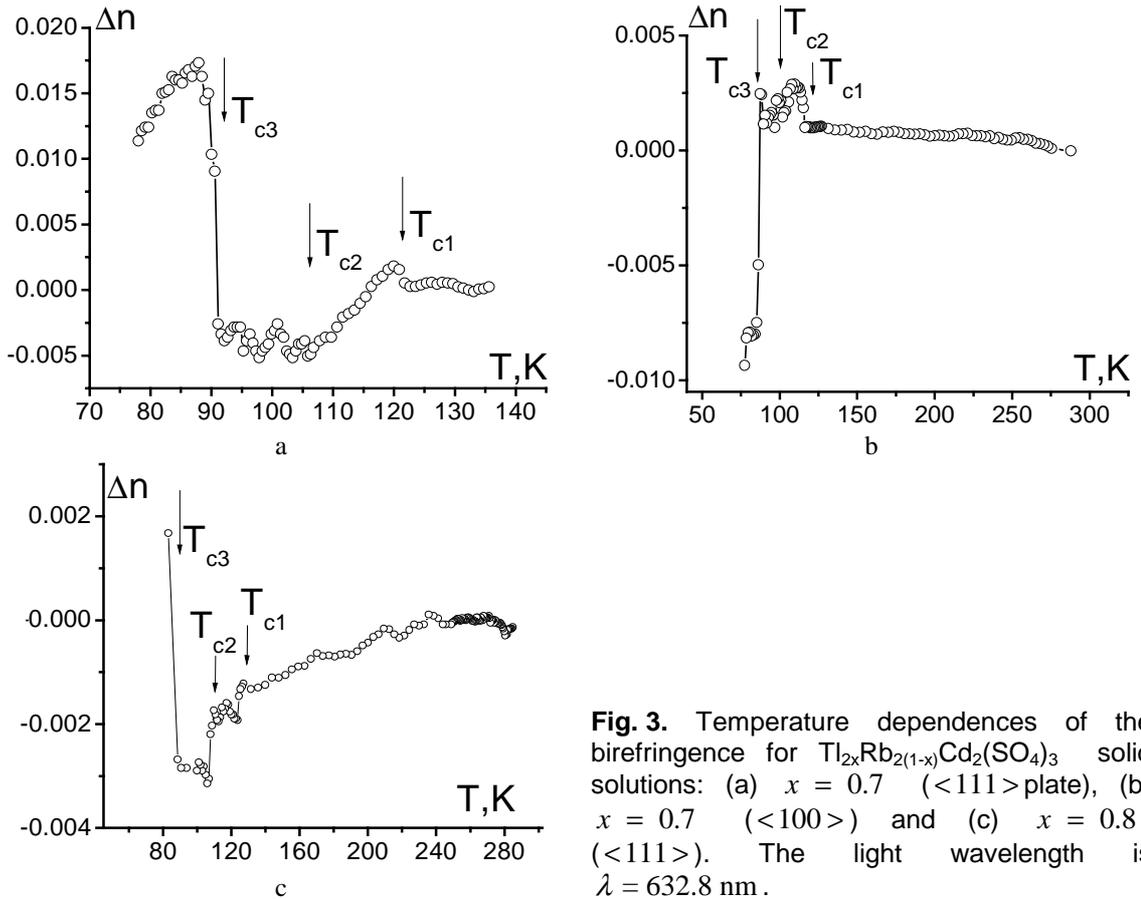

**Fig. 3.** Temperature dependences of the birefringence for $Tl_{2x}Rb_{2(1-x)}Cd_2(SO_4)_3$ solid solutions: (a) $x = 0.7$ (<111> plate), (b) $x = 0.7$ (<100>) and (c) $x = 0.8$ (<111>). The light wavelength is $\lambda = 632.8$ nm.

It is interesting to notice that the domain structure observed at the room temperature for the crystals with $x = 0.8$ (see Fig. 5) looks should have a residual character. A small value of the birefringence points to this (the samples placed in the "diagonal" position between the crossed polarizes are "grey"). Pinning of the impurities in the vicinity of walls inside the ferroic phase can cause availability of traces of the domain walls in the parent phase. For the mixed crystals, the substitution ions can play a role of such the impurities. Above the phase transition temperature, these impurities usually remain to be concentrated near their preceding positions.

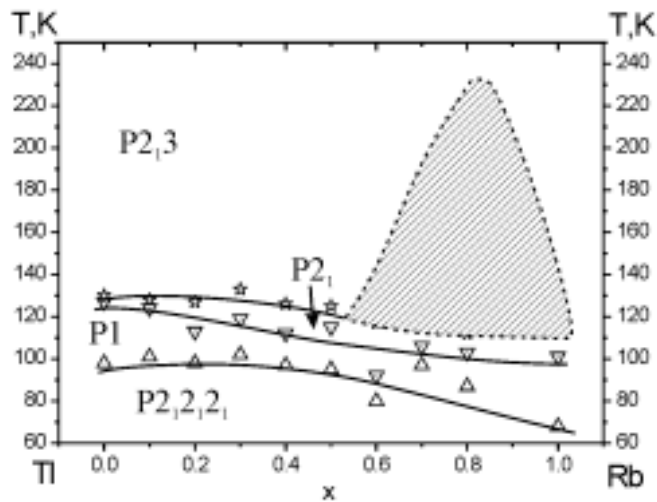

**Fig. 4.** $x,T$-phase diagram for the solid solutions of $Tl_{2x}Rb_{2(1-x)}Cd_2(SO_4)_3$ (the shaded region indicates where the $P2_13$ and $P2_1$ phases coexist).

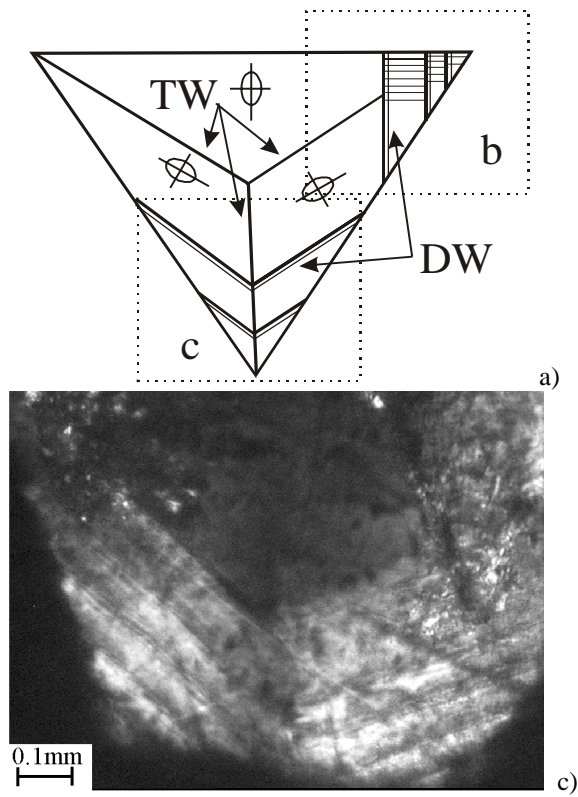
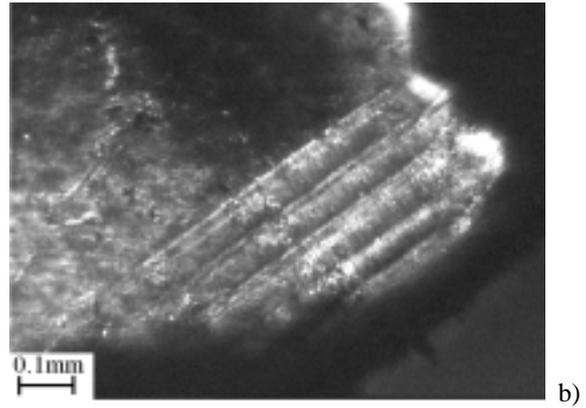
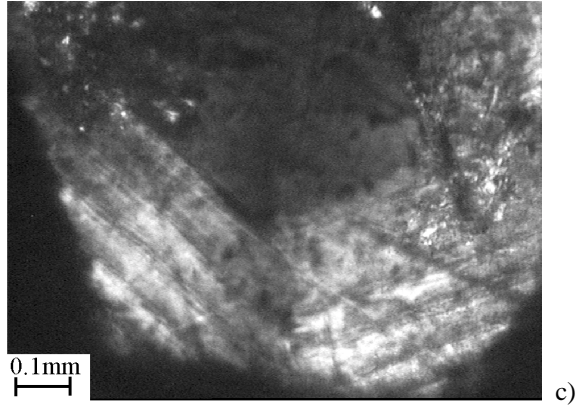

**Fig. 5.** Residual domain structure in $Tl_{2x}Rb_{2(1-x)}Cd_2(SO_4)_3$ solid solutions with $x = 0.8$ observed at the room temperature: (a) schematic view ($120°$ walls crossed in the central part are the twin walls – TW), (c) and (b) residual domain walls (DW) in different regions of sample.

While cooling the sample with $x = 0.8$, we have observed just an increase in the birefringence occurring in a wide temperature range that covers the monoclinic phase $P2_1$. New domains appear in the temperature range of the triclinic phase $P1$, coexisting simultaneously with the residual domain structure. It is only in the course of ferroelastic phase transition into $P2_12_12_1$ phase that the domain configuration changes drastically, accompanied with appearance of the domains separated by $120°$ domain walls, whose orientation coincides with that of the receding domain walls. After heating and passing the phase transition point $T_{c3}$, the residual domain structure disappears in the most of regions in the sample. Only $120°$ walls that cross all the sample do not disappear at both heating and cooling. Obviously, these walls should be twin walls. The extinction positions for the neighbouring twins differ by $30°$, but the extinction positions for the receding domains are the same. We have also observed the interference colours above $T_{c1}$ (approximately up to $220\,K$).

Thus, one can conclude that the residual domain structure observed at the room temperature corresponds to a so-called "forbidden" ferroelastic domain structure [15] appearing inside the $P2_12_12_1$ phase. The existence of interference colours up to $220\,K$ suggests that the phase transition at $T_{c1}$ has a diffuse character. Moreover, disappearance of the acoustic echo observed in our work [5] below $T = 220\,K$ at cooling is related to increasing anisotropy of crystals, disorientation of crystallographic axes in the twins and the residual domains, which lead to acoustic wave scattering. As a consequence, the temperatures above $T_{c1}$ in the crystals with $x = 0.7$, $0.8$ and 1.0 should represent a region where the $P2_13$ and $P2_1$ phases coexist rather than a region of a new phase with the symmetry $R3$.

## Conclusions

It is found that all of the as-grown solid solutions $Tl_{2x}Rb_{2(1-x)}Cd_2(SO_4)_3$ possess a residual birefringence at the room temperature. At the same time, the magnitude of the birefringence abruptly increases up to the value of $\Delta n = 6 \times 10^{-4}$, when $x$ is approaching $0.9$. Such a large birefringence cannot be caused solely by the imperfections introduced during the growing process. Using the polarization-microscopic observations, we have revealed a residual domain structure in the solid solutions $Tl_{2x}Rb_{2(1-x)}Cd_2(SO_4)_3$ ($x = 0.8$) at the room temperature. We have shown that the observed domain configuration corresponds to the remaining "forbidden" ferroelastic domain structure, which is preserved originating from the low-temperature ferroelastic phase. Basing on the studies for temperature variation of the birefringence and the observations of domain structure transformations, one can conclude that the phases with the symmetries $P2_13$ and $P2_1$ coexist above $T_{c1}$ in the compounds with $x = 0.7 - 1.0$. The phase transition temperatures $T_{c1}$, $T_{c2}$ and $T_{c3}$ obtained by us from the studies of temperature variations of the birefringence are in good agreement with those previously obtained on the basis of thermal expansion data and ultrasonic wave velocities.

## Acknowledgement

The authors are grateful to the Ministry of Education and Science of Ukraine (the Project N0103U000700) for financial support of the present studies.